# Scaling of Rainfall Intensity and Frequency with Rising Temperatures

Jun Yin[1], Bei Gao[2], Amilcare Porporato[3,4]

[1]Department of Hydrometeorology, Nanjing University of Information Science and Technology, Nanjing, 210044, China

[2]School of Environmental Science and Engineering, Nanjing University of Information Science and Technology, Nanjing, 210044, China

[3]Department of Civil and Environmental Engineering, Princeton University, Princeton, 08540, USA.

[4]High Meadows Environmental Institute, Princeton University, Princeton, 08540, USA

Global warming is projected to intensify the hydrological cycle, amplifying risks to ecosystems and society[1,2]. While extreme rainfall appears to exhibit stronger sensitivity to global warming compared to mean rainfall rates[3–5], a unifying physical mechanism capable of explaining this systematic divergence has remained elusive. Here, we integrate theory and data from a global network of nearly 50,000 rain-gauge stations to unravel the rainfall intensity and frequency response to rising temperatures. We show that the distributions of wet-day rainfall depth exhibit self-similar shapes across diverse geographical regions and time periods. Combined with the temperature response of rainfall frequency, this consistently links mean and extreme precipitation at both local and global scales. We find that the most probable change in rainfall intensity follows Clausius-Clapeyron (CC) scaling with variations shaped by a fundamental hydrological constraint. This behavior reflects a dynamic intensification of updrafts in space and time, which produces localized heavy precipitation events enhancing atmospheric moisture depletion and hydrologic losses through runoff and percolation. The resulting reduction in evaporative fluxes slows the replenishment of atmospheric moisture, giving rise to the observed trade-off between rainfall frequency and intensity. These robust scaling laws for rainfall shifts with temperature are essential for climate projection and adaptation planning.

Anthropogenic greenhouse gas emissions have altered the Earth's energy balance, leading to shifts in rainfall patterns[1,5]. Rainfall appears to be intensifying in warming climates[3,4], increasing the likelihood of extreme events. A key benchmark for understanding rainfall response to climate change is the Clausius-Clapeyron (CC) relationship, which predicts a ~7% increase in saturation vapor pressure per degree of warming. This is in line with the projected intensification of heavy precipitation at the global scale, as reported in the latest the Intergovernmental Panel on Climate Change (IPCC) assessment[5]. In contrast, global long-term mean precipitation is likely to increase at only 1–3% per degree of warming[5], making it increasingly difficult for land to balance the rising evaporative demand of a warmer atmosphere[6,7]. Such divergent trends result in complex hydrological patterns, causing both severe floods and prolonged droughts, thereby compounding water resource challenges and risks for ecosystems and society[2].

The divergent scaling in extreme and mean rates has been interpreted as signaling a reduction in rainfall frequency, a shift toward more skewed rainfall distributions, or a combination of both[1,8–11], with justifications based on both dynamics and thermodynamics[12–15]. Some studies posit that declining relative humidity over land reduces near-surface water vapor thus weakening the sensitivity of mean precipitation to warming[16,17], although this mechanism does not extend to oceans, where relative humidity changes are minimal[18,19]. Research has overwhelmingly focused on extreme events, often defined using high local percentiles, paying less attention to clearly separating the role of precipitation intensity to that of the event occurrence (i.e., rainfall frequency). Focusing on both rainfall intensity and frequency and using nearly 50,000 long-term rain gauge records worldwide, we find a complete chain of links in rainfall statistics that explains the divergent scaling of means and extremes from point to global scales. Through theoretical development, we demonstrate that the land-atmosphere water balance, operating within fundamental thermodynamic and hydrologic constraints, governs the response of rainfall frequency and intensity to global warming.

**Rainfall Scaling Relationships**

On account of rainfall's intermittent nature[20,21], the long-term rainfall rate can be expressed as the product of the long-term rainfall event frequency, $\lambda$, and the mean depth of rainfall events (also referred to as rainfall intensity in this study), $\bar{h}$,

$$\bar{P} = \lambda \bar{h}. \qquad (1)$$

The overbar denotes a temporal averaging operator over a long-term period (e.g., a 15-year interval), ensuring that the resulting statistics represent stationary climatological values. Seasonal partitioning of these statitstics (e.g., monthly averages) is not considered here, in order to maintain focus on extremes across mulitple years. At the daily timescale, $\lambda$ and $\bar{h}$ can be interpreted as the wet-day frequency and mean wet-day rainfall depth, where a wet day is defined as a day with accumulated rainfall depth exceeding a predefined threshold. This daily interpretation is adopted throughout, as 24 hours is the standard sampling interval for most rainfall observations.

The change of mean rainfall rate with temperature per unit rainfall rate can be linked to the related change with temperature of depth and frequency (see Methods),

$$\frac{d\bar{P}}{\bar{P}dT} = \frac{d\lambda}{\lambda dT} + \frac{d\bar{h}}{\bar{h}dT}, \qquad (2)$$

where, as for rainfall statistics, the temperature $T$ refers to the climatological values. To extend this relationship to extreme rainfall depths, we analyzed the Global Historical Climatology Network daily (GHCN-D) precipitation dataset[22], a global, quality-controlled collection of daily precipitation measurements from land-based weather stations worldwide (see Methods). Of the over 100,000 stations, nearly half of them have sufficiently long quality-controlled records (see supplementary Fig. S1). For these stations, we conducted a statistical analysis of the wet-day rainfall depth over a 15-year non-overlapping window to find its mean and percentiles (i.e., the daily precipitation extremes); sub-daily extremes (e.g., hourly intensity), while critical for engineering hydrology applications such as flash flood assessment[23,24], are not considered here, although they could be inferred from intensity-duration-frequency relationships[25,26].

As shown in Fig. 1a, the data reveal a generally a linear relationship between the mean wet-day rainfall depth $\bar{h}$ and its percentiles $h_{80}$ and $h_{90}$ no matter where and when these 15-year statistics are conducted. This is consistent with prior studies with assumption of exponential distribution of rainfall depth[27,28], although, as we explore later, this assumption can be relaxed to accommodate more complex forms of distributions. These linear relationships appear stationary, starting from the late 18th century when the earliest long-term rainfall records are available. This is confirmed by explicitly comparing the changes in mean rainfall depth and its percentiles between two 15-year periods in recent decades under the increasing warming trends. As shown in Fig. 1c, the changes in percentiles are proportional to the changes in the mean values. Moreover, these $\Delta\bar{h}$ - $\Delta h_{80}$ and $\Delta\bar{h}$ - $\Delta h_{90}$ slopes in Fig. 1c are close to those $\bar{h}$ - $h_{80}$ and $\bar{h}$ - $h_{90}$ slopes in Fig. 1a, further corroborating the assumption that the mean and percentiles move along these invariant linear lines under climate change.

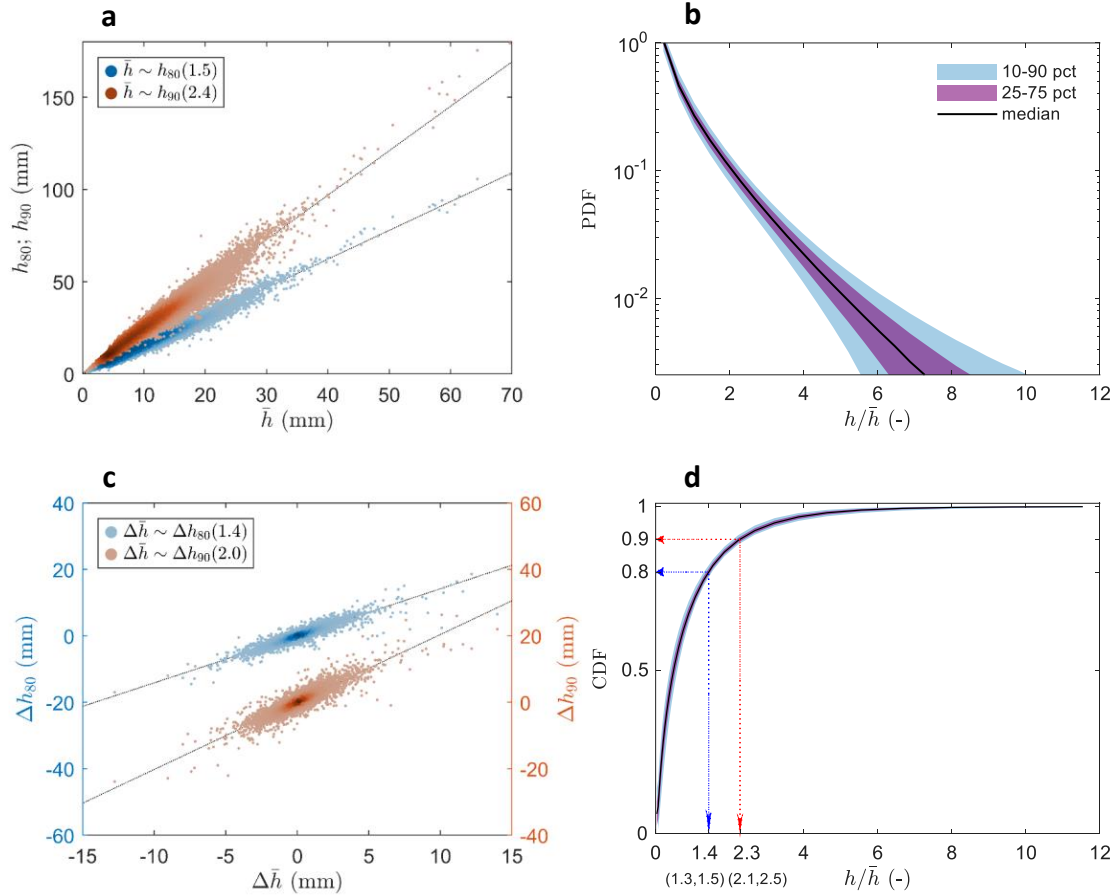

Fig. 1. Statistics of wet-day rainfall depths from global rain-gauge observations. Mean and percentiles of wet-day rainfall depths, denoted as $\bar{h}$ and $h_{80}$ or $h_{90}$, were estimated from GHCN-D rain-gauge records. Scatter points of mean and percentiles are presented in panel (**a**), in which each point refers to the paired statistics estimated from each site over a 15-year non-overlapping window; scatter plots of their changes, denoted as $\Delta\bar{h}$ and $\Delta h_{80}$ or $\Delta h_{90}$, between 1995-2009 and 2010-2024 are presented in panel (**c**). The dotted lines represent linear fits with corresponding slopes annotated in each panel's legend. The color of the dots corresponds to the local density with darker hues indicating higher concentrations. The distributions of normalized rainfall depth, $h/\bar{h}$, were estimated for each site over a 15-year non-overlapping window and the spread of these (**b**) PDFs and (**d**) CDFs are presented in shaded areas. The arrows in (d) point the median and the spread of CDF at two specific values, corresponding to $h_{80}$ and $h_{90}$.

This linearity extends to any percentile of rainfall depth (see supplementary Fig. S2 for additional results from the 10th, 20th, …, 70th percentiles), although it exhibits relatively higher dispersion for lower percentiles. This implies that the distribution has scalable properties and maintains a self-similar shape. To further test this, we stacked the probability density functions (PDFs) of normalized rainfall depth from all GHCN-D stations across non-overlapping 15-year windows. The normalized depth shown in Fig. 1b consistently converges toward a unique shape characterized by a short power-law head followed by a long exponential tail. Such a pattern is consistent with the gamma distribution derived analytically from the relationship between column water vapor and precipitation[29]. The observed self-similarity of normalized distributions captures much of the variability across rainfall depth distributions, which have historically been represented using a wide range of mathematical forms of varying complexity in hydrological studies[30–33] and in analyses of climate change impacts on rainfall distributions[34–38]. It also aligns with the "shift mode" or linear scaling feature documented in prior studies[39–41]. Importantly, however, no specific functional form is assumed in our analysis, and the results remain robust regardless of the chosen distribution, provided that the distribution itself is self-similar.

The corresponding cumulative distribution functions (CDFs) is shown in Fig. 1d. The percentile of rainfall depth is proportional to the mean depth,

$$h_a = k_a \bar{h},  \qquad (3)$$

where the subscript refers to the percentile and $k_a$ is the inverse function of the CDF. This linearity indicates the same relative changes for both mean and percentiles, i.e., $dh_a/(h_a dT) = d\bar{h}/(\bar{h} dT)$. It further allows us to generalize Eq. (2) as (see Methods)

$$\frac{d\bar{P}}{\bar{P} dT} = \frac{dh_a}{h_a dT} + \frac{d\lambda}{\lambda dT},  \qquad (4)$$

thereby establishing a direct link between the temperature scaling of mean rainfall rate and the rainfall depth at any intensity, including extreme events. Equation (4) represents a major result of this study, as it shows that the scaling behavior of rainfall extremes can be traced back to variations in wet-day frequency.

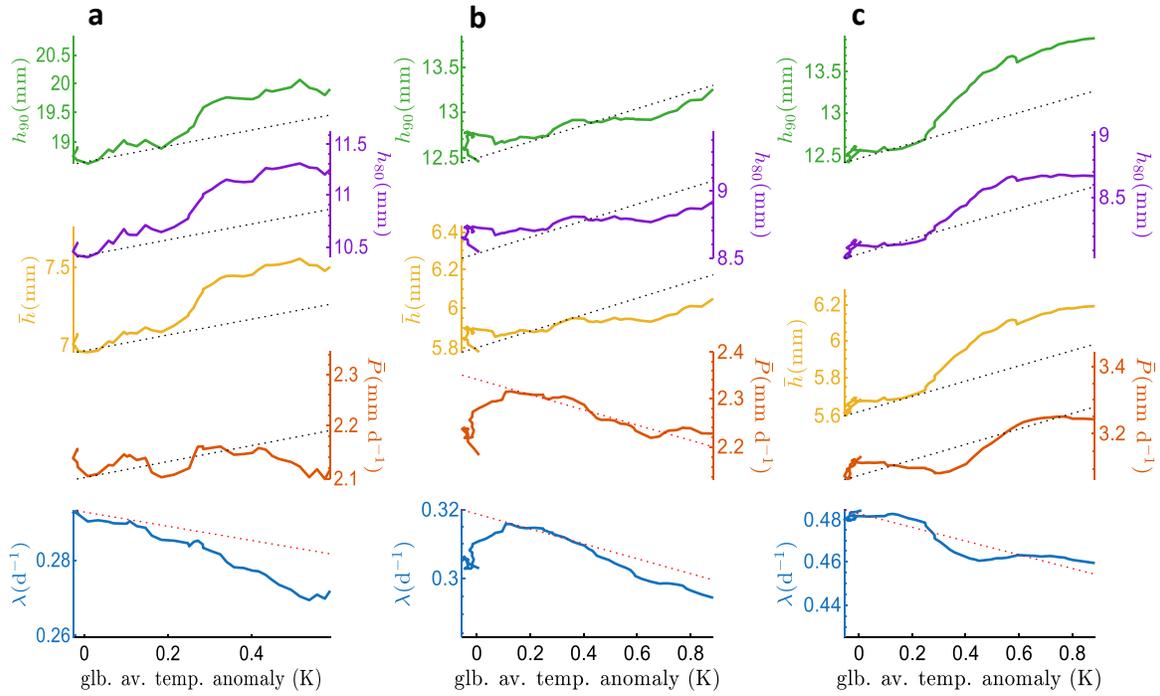

Fig. 2. Precipitation scaling relationships at regional and global scales. Wet-day frequency, $\lambda$, mean wet-day rainfall depth, $\bar{h}$, and its percentiles, $h_a$, spatially averaged over (a) East Asia, (b) global land, and (c) global ocean are plotted against global mean temperature. Rainfall statistics from (a) and (b, c) are estimated from GHCN-D and ERA-5, respectively; global temperature anomalies are from GISTEMP v4. The black and red dotted lines refer to positive and negative CC scaling rates; all y-axes are logarithmically scaled, though the narrow ranges give them quasi-linear appearances. Rainfall records from GHCN-D after 2000 in East Asia were excluded from analysis due to incomplete data (i.e., more than 10% combined missing or invalid values). More results for other regions from GHCN-D are reported in supplementary Fig. S4.

## Global Validation

We validated the scaling relationships across spatially aggregated regional and global scales. We again used the GHCN-D dataset and focused on reference regions of the IPCC special report on extreme events[42] (see supplementary Fig. S3). We adopted 15-year moving windows to provide smooth variations of rainfall statistics. Station statistics were then spatially averaged with equal weight for each station over each reference domain to reduce the impacts of climate variability and then compared against global mean temperature anomaly (Fig. 2). In East Asia (Fig. 2a), the mean rainfall rate shows large climate variability but negligible trends. In contrast, the mean wet-day rainfall depth significantly increases under warmer climates at CC or sometimes super CC rates, aligning with previous studies[37,43–45]. These gaps are simply filled by the negative temperature scaling of wet-day frequency as theoretically proved in Eq. (2). As seen globally (see supplementary Fig. S4), a negative wet-day frequency scaling tends to amplify rainfall depth scaling, while a positive wet-day frequency scaling attenuates it. Moreover, the nearly parallel curves of $\bar{h}$, $h_{80}$, and $h_{90}$ on the logarithmic scale of Fig. 2a and in supplementary Fig. S4 reveal a similar scaling of mean and extreme rainfall depths, corroborating the theoretical relationship in Eq. (3).

We followed the same procedure to assess the validity of the scaling relationship globally, using the fifth-generation reanalysis from the European Centre for Medium-Range Weather Forecasts (ERA5)[46], a global precipitation dataset at 0.25-degree grid resolution widely used for analysis of climate trends and extreme events. In agreement with prior studies[47,48], a reduction in wet-day frequency has been observed across most regions. Over the land (Fig. 2b), the decrease of rainfall frequency is accompanied by a proportional, yet slightly slower reduction in mean rainfall rate, leading to modest increase in rainfall depth. Over the ocean (Fig. 2c), the decline of wet-day frequency is concurrent with a rise in mean rainfall rate, resulting in a super-CC scaling of rainfall depth in certain temperature range.

It is also important to notice that the scaling becomes muddled if the extremes are defined, as in many previous studies, using high percentiles of rainfall rates across all days, including a large fraction of dry days with zero rainfall (our study ranks percentiles based only on wet-day rainfall depth). For example, the 90th percentile of wet-day rainfall depth ($h_{90}$) in a region with a typical wet-day frequency of 0.2 corresponds approximately to the 98th percentile of rainfall rate ($P_{98}$). To examine this effect, we repeated the analysis using different rainfall rate percentiles. As shown in supplementary Fig. S5, when rainfall rates are very extreme (e.g., $P_{99}$), their scaling rates closely match those of mean or extreme rainfall depth (e.g., $\bar{h}$ or $h_{90}$); whereas for less extreme rainfall rates (e.g., $P_{85}$), the scaling rates approach those of mean rainfall rate (i.e., $\bar{P}$). Similar patterns have been reported in independent studies, which also found smaller scaling rates for less extreme rainfall (e.g., Fig. 5 in reference[49]). Together, these results highlight that variations in dry-day frequency (i.e., the prevalence of zeros) exert little influence on the scaling of very extreme rainfall rates but become increasingly important for less extreme rainfall rates. This conclusion is further corroborated by theoretical analyses of extremes with and without the inclusion of zeros (see Methods).

**Thermo-Hydrologic Constraints on the Rainfall-Variability Distribution**

The distribution of rainfall changes with temperature across different locations reveals key thermodynamic and hydrologic constraints, which are essential for informing models of future rainfall scenarios. As a starting point, consider that the additive form of Eq. (4) establishes a direct relationship between the relative changes in mean (or extreme) rainfall depth, frequency, and mean rate. This relationship is visualized in the four-quadrant diagram of Fig. 3a, categorizing the relative changes in frequency and depth. The first and third quadrants denote overall wetting and drying trends, respectively, as the frequency and depth change in the same direction. In contrast, the second quadrant reflects a more intermittent rainfall regime (more intense but less frequent), whereas the fourth indicates a more regular regime (less intense but more frequent). This follows from the fact that iso-lines with a slope of -1 correspond to constant changes in mean rainfall rate, with compensating variations in rainfall frequency and depth. The joint distributions of the fitted slopes of rainfall depth and frequency against global mean temperature from GHCN-D stations at different stations are plotted as a contour map in this plane in Fig. 3a, along with their geographic location shown in Fig. 3b (also see supplementary Table S1). The majority exhibit trends toward wetter regimes, whereas drying trends are more common only in Australia and West North America (see supplementary Table S2). Moreover, stations predominantly fall within the second and fourth quadrants, indicating shifts toward either more intermittent (e.g., East Asia, North Asia, and Central America/Mexico) or more regular (e.g., Australia and West/East North America) rainfall regimes, as indicated by the first principal component of the local rainfall frequency and depth scaling distribution. This anticorrelation remains visible in the precipitation scaling with local temperature (see supplementary Fig. S6) but is lost in the ERA5 reanalysis data at global coverage (see supplementary Fig. S7). When sampled over the geophysical locations of the GHCN-D stations, the corresponding ERA5 results show a second mode falling on the second quadrant (see supplementary Fig. S8), suggesting that the anticorrelation between frequency and depth scaling has only been partially captured in the data assimilation.

The thermodynamic CC scaling, governing the moisture content in the low-level atmosphere in wet events, is also plotted in Fig. 4a. While most stations exhibit a sub-CC scaling (see supplementary Table S3 and Fig. S9), the mode of this joint distribution falls precisely on the CC scaling line. This is interesting as it shows that the most likely change in rainfall regimes is set by this thermodynamic constraint. A similar result is also found for the ERA5 reanalysis data (see supplementary Fig. S7 and Fig. S8). The mode falls slightly below the CC scaling line when rainfall depth and frequency are regressed against local temperature (see supplementary Fig. S6). Lower scaling rates for local temperatures are consistent with the more rapid warming over land compared to the global mean temperature.

Also shown, as a blue dashed line with a -1 slope, is the global energetic constraint—estimated at roughly 2.5% $K^{-1}$—under which changes in latent heat release from precipitation balance atmospheric cooling, yielding relatively small changes in the mean rainfall rate[14]. This constraint reflects the global, long-term balance between precipitation and evaporation, but it does not account for local hydrologic limitations over land, where precipitation must partition into evapotranspiration, surface runoff, and percolation. At local scale, this influences atmospheric dynamics, altering moisture convergence and horizontal energy transport[12,50] and thereby producing deviations from the global energetic limits.

These local constraints are represented by a minimalist model of the terrestrial water balance (see Methods), whose solution closely aligns with the first principal axis (see green dash-dot line in Fig. 3a). When extended to global conditions, it coincides with the -1 slope noted earlier. This constraint implies dynamic adjustments of updrafts in space and time, which can enhance precipitation efficiency and produce departures from the CC scaling of rainfall depth. Regions that experience deep removal of atmospheric moisture during super-CC events may require longer recovery times before the next wet event, leading to lower rainfall frequency. Intense rainfall over land also increases runoff generation, leaving less water available for evaporation and slowing the atmospheric moisture recharge, further reducing rain-event frequency. Conversely, weakened updrafts and lower precipitation efficiency yield sub-CC scaling of rainfall depth. Regions with more regular rainfall tend to have smaller runoff coefficients, allowing a larger share of precipitation to be returned to the atmosphere via evaporation. Inefficient removal of atmospheric moisture and higher evaporative fluxes accelerate the recovery of precipitable water, promoting more frequent rainfall. In general, more intermittent rainfall loses a greater fraction of water as runoff, whereas more regular rainfall recycles more water back into the atmosphere. This also explains why the increase in mean rainfall rate is relatively smaller in the second quadrant than in the fourth, consistent with the small yet distinct angle between the local hydrological-constraint and the global energetic-constraint lines shown in Fig. 3a.

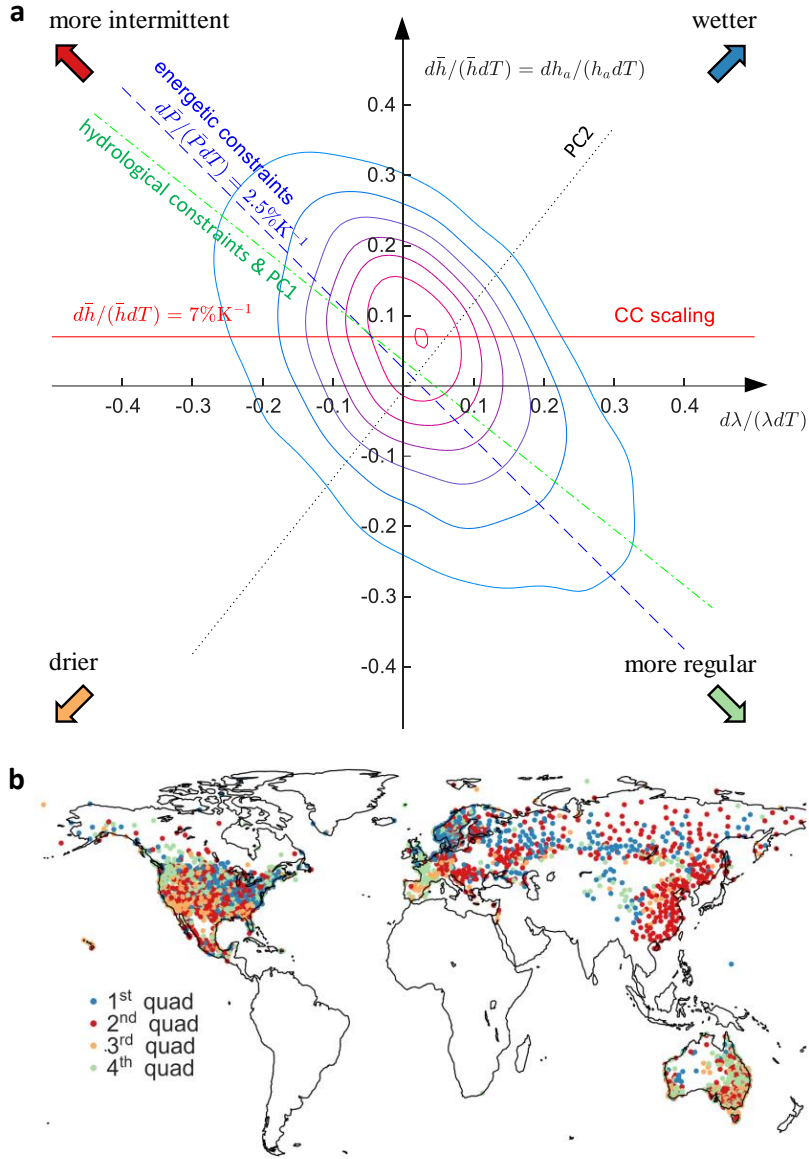

Fig. 3. Distribution of local rainfall scaling, thermodynamic, and hydrologic constraints. (**a**) Diagram with x-axis and y-axis refering to relative changes in wet-day frequency and mean depth (or its percentiles), respectively. Lines with slope of -1 have the same relative changes in total precipitation rate. The blue dash line refers to the global energetic constraint with relative changes in mean precipitation at 2.5% K$^{-1}$ estimated from statistical-equilibrium simulations[14]; the red horizontal line marks the CC scaling rate of 7% K$^{-1}$. The contour shows the joint distribution of the relative changes in local rainfall frequency and depth from GHCN-D records with red hue indicating higher probability. The green dash-dot line refers to the hydrological constraint and also the first principal component (PC1); the black dot line is the second principal component (PC2) (**b**) Geographical location of GHCN-D stations falling on each quadrant of the diagram in (a). Note only stations with continous records after 1960 are analyzed (see Methods).

**Conclusions**

The self-similar form of rainfall–depth distributions, together with a clean separation between precipitation frequency and intensity, offers a coherent route for incorporating temperature effects into rainfall-regime projections. Building on established statistical and physical links between column water content and wet-day rainfall depth[51–53], hydrologic and thermodynamic constraints on intensity yield a global temperature scaling for extreme rainfall. This framework also supports stochastic downscaling of future rainfall scenarios by using relatively robust projections of mean rainfall rates and translating them into full distributional behavior, which offers an advantage for climate-risk assessment.

By neglecting rainfall frequency, prior work has often implied that super-CC scaling is mainly an extreme-event phenomenon. Centrally accounting for frequency indicates that super-CC behavior can arise across rainfall depths through an intensity–frequency trade-off: intensification can be amplified by concurrent declines in event frequency. Explicit treatment of frequency helps reconcile divergent trends in mean rainfall when dry days are included, and links the commonly observed decline in rainfall frequency with warming to shifts in aridity and ecohydrological dynamics[21,54]. This coupling between the intensity-frequency trade-off and land-atmosphere water-balance constraints provides a physically grounded basis for projecting future changes in rainfall variability.

## Methods

### Precipitation scaling relationships

As given in Eq. (1) in the main text, the long-term mean rainfall rate can be expressed as the multiplication of rainfall event frequency, $\lambda$, and mean depth of each rainfall event, $\bar{h}$. The derivative of Eq. (1) with respect to temperature, $T$, is,

$$\frac{d\bar{P}}{dT} = \lambda \frac{d\bar{h}}{dT} + \bar{h}\frac{d\lambda}{dT}. \qquad (5)$$

Dividing Eq. (5) by (1) yields Eq. (2) in the main text. It presents the scaling relationships between mean rainfall rate, depth, and frequency.

The derivative of Eq. (3) in the main text is

$$\frac{dh_a}{dT} = k_a \frac{d\bar{h}}{dT}, \qquad (6)$$

Dividing Eq. (6) by (3) yields the scaling relationships between mean and percentiles of rainfall depth, i.e.,

$$\frac{dh_a}{h_a dT} = \frac{d\bar{h}}{\bar{h} dT}. \qquad (7)$$

Combination of Eqs. (2) and (7) gives Eq. (4) in the main text, which suggests that the differences between the scaling of mean rainfall rate and rainfall depth of any percentiles are accounted for by the variations of wet-day frequency.

### Extremes with and without counting zero-records

Conventionally, quantifying percentiles of daily rainfall rates, $P_b$, needs to rank all daily rainfall from zero rainfall records to the maximum values; quantifying rainfall depth percentiles, $h_a$, in this study only counts the wet-day rainfall depths. To understand their scaling relationships, we numerically equate these two percentiles and use Eq. (3)

$$P_b = h_a = k_a \bar{h}, \qquad (8)$$

which requires that[55]

$$\frac{a}{100} = 1 - \frac{1 - \frac{b}{100}}{\lambda}, \tag{9}$$

so that *a* becomes smaller than *b* to account for all these 0's records.

To investigate the temperature scaling of $P_b$, we fix *b* and let the *a* vary according to Eq. (9). The derivative form of Eq. (8) can be expressed as

$$\frac{dP_b}{P_b dT} = \frac{dh_a}{h_a dT} = \frac{d\overline{h}}{\overline{h}dT} + \frac{dk_a}{k_a dT}, \tag{10}$$

which is different from Eq. (6) since *a* and $k_a$ are now variables instead of constants.

The last term $dk_a / (k_a dT)$ in Eq. (10) accounts for different scaling of rainfall rate percentile and mean rainfall depth. To estimate this term, we approximate the CDF in Fig. 1d as an exponential distribution with unit mean

$$\frac{a(k)}{100} = 1 - e^{-k}, \tag{11}$$

The inverse function is

$$k = \ln\left(\frac{100}{100 - a}\right). \tag{12}$$

With this $k_a$ function and Eq. (9), we have,

$$\frac{dk_a}{k_a dT} = \frac{1}{k_a \lambda} \frac{d\lambda}{dT}. \tag{13}$$

Substitution Eq. (13) into Eq. (10) yields

$$\frac{dP_b}{P_b dT} = \frac{d\overline{h}}{\overline{h}dT} + \frac{d\lambda}{k_a \lambda dT}. \tag{14}$$

It suggests that the different scaling of rainfall rate percentile and mean rainfall depth is linked to the variations of wet-day frequency.

To explain these scaling relationships, we consider an example in a region with typical wet-day frequency 0.25. When considering the very extreme rainfall, e.g, $P_{99}$, the corresponding $k_a$ (= 3.2) is relatively large. Therefore, the last term of Eq. (14) is relatively small and the scaling of very-extreme rainfall is close to that of mean rainfall depth (i.e., $dP_b / (P_b dT) \approx d\bar{h} / (\bar{h} dT)$, for large $a$). When considering less-extreme rainfall, e.g., $P_{90}$, the corresponding $k_a$ is close to 1. Comparison between Eqs. (14) and (2) suggests that the scaling of less-extreme rainfall is close to that of mean rainfall rate (i.e., $dP_b / (P_b dT) \approx d\bar{P} / (\bar{P} dT)$, for moderate $a$). This corroborates the statistical analysis in supplementary Fig. S5.

**Minimalist approach to land-atmosphere water balance**

The long-term water balance of the coupled atmospheric column and the land surface can be expressed as[56]

$$\bar{P} = \bar{E} + \overline{LQ} = \bar{E} + \bar{A}, \tag{15}$$

where $P$ is precipitation, $E$ is evaporation, LQ is runoff and deep percolation, and $A$ is atmospheric moisture convergence; this implies that the long-term averages of atmospheric convergence and surface runoff and percolation must be equal, $\overline{LQ} = \bar{A}$.

The atmospheric moisture balance is maintained by continuous replenishment from evaporation and advection, against intermittent removal from precipitation. Combining Eq. (1) and (15), the rainfall frequency can be also written as

$$\lambda = \frac{\bar{E} + \bar{A}}{\bar{h}}, \tag{16}$$

simultaneously incorporating both land and atmosphere water balances.

To relate rainfall frequency and depth to surface and atmospheric processes, we use a minimalist hydrological model[21,57] (also see supplementary Text 1.1), in which rainfall is a stochastic input with explicit frequency and depth representation, and evaporation is proportional to potential evaporation, $E_{max}$, and relative soil moisture, $s$. The model can be analytically solved to express long-term average evaporation as a function of rainfall frequency, depth, potential evapotranspiration, and soil water storage capacity ($w_0$). Combined with Eq. (16), this analytical solution enables quantification of the relationship between mean rainfall frequency and depth as shown in Fig. S10. Increases in mean rainfall depth correspond to decreases in rainfall frequency, giving rise to the observed anti-correlated pattern in a manner that is robust to changes in hydrological parameters.

To link the rainfall solution to temperature changes, we set typical hydrologic conditions and specify the temperature sensitivities of $E_{max}$ and $\overline{LQ}$ (or $\bar{A}$). Under a small temperature perturbation, the hydrological model predicts deviations in both rainfall frequency and intensity from their baseline values, allowing us to derive their scaling relationships. With typical values ($\lambda = 0.3$, $\bar{h} = 15$ mm, $w_0 = 120$ mm, $E_{max} = 6$ mm d$^{-1}$, $\bar{A} = 1$ mm d$^{-1}$, $dE_{max}/(E_{max}dT) = 4\%$ K$^{-1}$, $d\bar{A}/(\bar{A}dT) = 4\%$ K$^{-1}$), the resulting curve is indistinguishable from the first principal component in Fig. 3a of the main text (referred to as the hydrological constraint).

At the global scale, lateral flow is not present ($A = 0$), and surface evaporation is at its maximum rate, being dominated by ocean contribution ($E = E_{max}$). This means that Eq. (16) becomes

$$E_{max} = \lambda \bar{h} , \qquad (17)$$

so that its specific derivative is

$$\frac{dE_{max}}{E_{max}dT} = \frac{d\lambda}{\lambda dT} + \frac{d\bar{h}}{\bar{h}dT} . \qquad (18)$$

This corresponds to a line of slope of -1 in Fig. 3a and is equivalent to the global energetic constraint.

**Statistical Analysis**

To estimate rainfall statistics from GHCN-D, we segmented rainfall records from each rain-gauge station into consecutive 15-year non-overlapping blocks. Data blocks were excluded if the combined missing values and quality-control failures surpasses 10% of potential observations. For each data block, we estimated the mean and percentiles of rainfall rate, $\bar{P}$ and $P_b$, from all those valid records. We define wet days as those with rainfall rate larger than 0.1 mm d$^{-1}$ (see supplementary text 1.2 for the choice of the threshold). Wet-day frequency, $\lambda$, was estimated as the ratio of wet-day number to the total number of valid observation days. Mean and percentiles of wet-day rainfall depth, $\bar{h}$ and $h_a$, were calculated exclusively from wet-day observations. Rainfall statistics from each station within each 15-year data block were plotted in Fig. 1a in the main text. We also estimated the changes in rainfall statistics between 1995-2009 and 2010-2024 as shown in Fig. 1c. Empirical PDF and CDF of normalized wet-day rainfall depth, $h/\bar{h}$, were estimated for each station within each 15-year block using the same bin spacing (see supplementary text 1.3 for the choice of bins). The spread (10-90 percentiles, 25-75 percentiles, and median) of all the PDFs and CDFs within each bin were presented in Fig. 1 b and d.

To estimate precipitation scaling relationships from GHCN-D, we performed the same rainfall statistics for each station but with 15-year moving windows. We calculated global mean near-surface temperature within these 15-year windows from Goddard Institute for Space Studies (GISS) Surface Temperature Analysis version 4[58]. The rainfall statistics over the reference regions developed for IPCC special report on extreme events were spatially averaged and plotted against the global mean temperature in Fig. 2a and supplementary Fig. S4.

To estimate precipitation scaling relationships from ERA5 datasets, we also performed the same rainfall statistics for grid but with wet-day threshold of 1 mm d$^{-1}$ (see supplementary text 1.2 for the choice of the threshold). These statistics in each grid were then spatially averaged over the land and ocean with geodetic weights measuring the relative size of each grid. These averages were plotted against the global mean temperature from GISS Surface Temperature Analysis version 4 in Fig. 2 b and c.

Gridded temperature at global coverage from GISS version 4[58] was also interpolated onto the GHCN-D rain-gauge locations and ERA5 grids and then used for analyzing the precipitation scaling against local temperature.

All rainfall statistics in the main text were estimated over 15-year window to reduce the impacts of climate variability. Similar patterns can be obtained for statistics over different lengths of periods (e.g., 10 years or 20 years, see supplementary Fig. S11).

**Data availability**

GHCN-D precipitation dataset is available at https://www.ncei.noaa.gov/products/land-based-station/global-historical-climatology-network-daily; ERA5 post-processed daily statistics of precipitation dataset can be downloaded from https://cds.climate.copernicus.eu/datasets/derived-era5-single-levels-daily-statistics?tab=overview; GISS Surface Temperature Analysis version 4 (GISTEMP v4) is available at https://data.giss.nasa.gov/gistemp/.

**Code availability**

Codes for data processing, percentile analyses, and minimalist hydrological models are available at github (https://github.com/jy-junyin/prcpscaling).

**Acknowledgment**

J. Y. and B.G. acknowledges support from the National Natural Science Foundation of China (42330604). A. P. acknowledges support from the Carbon Mitigation Initiative at Princeton.

**Competing interests**

The authors declare no competing interests.